\documentclass[fleqn,usenatbib]{mnras}

\usepackage{newtxtext,newtxmath}
\usepackage[T1]{fontenc}
\usepackage{color}

\DeclareRobustCommand{\VAN}[3]{#2}
\let\VANthebibliography\thebibliography
\def\thebibliography{\DeclareRobustCommand{\VAN}[3]{##3}\VANthebibliography}

\newcommand{\dmfif}{$\Delta m_{15,B}$}
\usepackage{graphicx}	% Including figure files
\usepackage{amsmath}	% Advanced maths commands

\title[SN~Encore Spectra]{Spectroscopic analysis of the strongly lensed SN~Encore: Constraints on cosmic evolution of Type Ia supernovae}

\author[S. Dhawan et al.]{S. Dhawan,$^{1}$\thanks{E-mail: sd919@cam.ac.uk}
J.~D.~R.~Pierel$^{2}$\thanks{NASA Einstein Fellow}, M.~Gu$^{3}$, A.~B.~Newman$^{4}$, C. Larison$^{5}$, M. Siebert$^{6}$, T.~Petrushevska$^{7}$, \newauthor
F.~Poidevin$^{8,9}$, S.~W.~Jha$^{5}$, W.~Chen$^{10}$, Richard~S. Ellis$^{11}$, B.~Frye$^{12}$, J.~Hjorth$^{13}$,Anton M. Koekemoer$^{2}$,  \newauthor I. P\'{e}rez-Fournon$^{8,9}$, A.~Rest$^{2,14}$,  T.~Treu$^{15}$,  R.~A.~Windhorst$^{16}$, Y.~Zenati$^{14,2}$
\\
$^{1}$Institute of Astronomy and Kavli Institute for Cosmology, University of Cambridge, Madingley Road, Cambridge CB3 0HA, UK\\
$^{2}$ Space Telescope Science Institute, Baltimore, MD 21218\\
$^{3}$ Department of Physics, The University of Hong Kong, Pok Fu Lam, Hong Kong\\
$^{4}$ Observatories of the Carnegie Institution for Science, 813 Santa Barbara St., Pasadena, CA 91101, USA\\
$^{5}$ Department of Physics \& Astronomy, Rutgers, State University of New Jersey, 136 Frelinghuysen Road, Piscataway, NJ 08854, USA\\
$^{8}$Instituto de Astrof\'{\i}sica de Canarias, V\'{\i}a   L\'actea, 38205 La Laguna, Tenerife, Spain\\
$^{9}$Universidad de La Laguna, Departamento de Astrof\'{\i}sica,  38206 La Laguna, Tenerife, Spain\\
$^{10}$ Department of Physics, Oklahoma State University, 145 Physical Sciences Bldg, Stillwater, OK 74078, USA\\
$^{11}$ Department of Physics \& Astronomy, University College London, London WC1E 6BT\\ 
$^{12}$ Department of Astronomy/Steward Observatory, 933 Cherry Avenue, University of Arizona, Tucson, AZ 85721 \\
$^{13}$ DARK, Niels Bohr Institute, University of Copenhagen, Jagtvej 155A, 2200 Copenhagen, Denmark \\
$^{14}$ Physics and Astronomy Department, Johns Hopkins University, Baltimore, MD 21218, USA\\
$^{15}$ Physics and Astronomy Department, University of California, Los Angeles CA 90095 \\
$^{16}$ School of Earth and Space Exploration, Arizona State University, Tempe, AZ 85287-1404, USA \\ 
}

\date{Accepted XXX. Received YYY; in original form ZZZ}

\pubyear{2015}

\begin{document}
\label{firstpage}
\pagerange{\pageref{firstpage}--\pageref{lastpage}}
\maketitle

\begin{abstract}
Strong gravitational lensing magnifies the light from a background source, allowing us to study these sources in detail. Here, we study the spectra of a $z = 1.95$ lensed Type Ia supernova SN~Encore for its brightest Image A, taken 39 days apart. We infer the spectral age with template matching using the supernova identification (\texttt{SNID}) software and find the spectra to be at 29.0 $\pm 5.0$ and 37.4 $\pm 2.8$ rest-frame days post maximum respectively, consistent with separation in the observer frame after accounting for time-dilation.  Since SNe~Ia measure dark energy properties by providing relative distances between low- and high-$z$ SNe, it is important to test for evolution of spectroscopic properties. Comparing the spectra to composite low-$z$ SN~Ia spectra, we find strong evidence for similarity between the local sample of SN~Encore. The line velocities of common SN~Ia spectral lines, Si II 6355 $\AA$ and Ca II NIR triplet are consistent with the distribution for the low-$z$ sample as well as other lensed SNe~Ia, e.g. iPTF16geu ($z = 0.409$)and SN~H0pe ($z = 1.78$).  The consistency in SN~Ia spectra across cosmic time demonstrates the utility of using SNe~Ia in the very high-$z$ universe for dark energy inference. We also find that the spectra of SN~Encore match the predictions for explosion models very well. With future large samples of lensed SNe~Ia, spectra at such late phases will be important to distinguish between different explosion scenarios.

%Type Ia supernovae are excellent distance indicators for cosmology. Strong gravitational lensing allows us to observe high-redshift SNe~Ia in detail. Comparing high- and low-$z$ SNe~Ia allows us to constrain the evolution of intrinsic SN~Ia properties, which is important for minimising systematics in inferring dark energy properties. We present a spectroscopic analysis of SN~Encore, an SN~Ia at $z = 1.946$ lensed by the foreground MACSJ0138 cluster. Based on a post-maximum light spectrum, we compare to the median low-$z$ spectra from the Carnegie Supernova Project and find no evidence from deviations from the mean spectra at the 1-$\sigma$ level. We also measure the line velocities and pseudo equivalent widths for the Si II and Ca II features and find no evidence for deviation from the distribution at low-$z$. \sdt{v old version: needs revision} %Confronting the observations with prediction from models for spectroscopically normal SNe~Ia, we find that both Chandrasekhar mass models and high primary white dwarf mass sub-Chandrasekhar mass models 
\end{abstract}

\begin{keywords}
supernovae: general -- supernovae:individual:SN~Encore
\end{keywords}

%%%%%%%%%%%%%%%%%%%%%%%%%%%%%%%%%%%%%%%%%%%%%%%%%%

%%%%%%%%%%%%%%%%% BODY OF PAPER %%%%%%%%%%%%%%%%%%

\section{Introduction}

Type Ia supernovae (SNe~Ia) are thermonuclear explosions of carbon-oxygen white dwarfs in a binary system \citep[see][for a review of the likely progenitor systems of SNe~Ia]{Maguire2017,Livio2018}. Their bright peak luminosity and largely uniform observable characteristics across the population makes them excellent distance indicators. SNe~Ia have been instrumental in the discovery of accelerated expansion \citep{Riess:1998cb,Perlmutter:1998np} and for precision estimates of dark energy properties \citet{Brout2022,DES2024,DESI2024} as well as the present day expansion rate i.e. the Hubble Constant \citet{Riess2022}. 
In optical wavelengths, the regime where most constraints on cosmology from SNe~Ia are obtained, standardisation of their peak luminosity can reduce the scatter to $\sim 15\%$. To obtain such a small scatter, the peak brightness is corrected for correlations with the lightcurve width and colour \citep[e.g.,][]{phillips1993,tripp1998} and also host galaxy properties \citep[e.g.,][]{Kelly2010,Sullivan2010,Ginolin2024a,Ginolin2024b}.

Measuring dark energy properties with SNe~Ia requires relative distance measurements, comparing the stretch and colour corrected brightness of the high-$z$ SNe~Ia with a low-$z$ ``anchor" sample \citep[e.g.,][]{Brout2022,DES2024}. Near-future experiments to study dark energy will sizably increase the sample of SNe~Ia at very high-$z$ - i.e. $z \gtrsim 2$ \citep{Hounsell2018}. With such large samples, the SN~Ia constraints on dark energy will be dominated by systematic uncertainties. One possible significant source of error is the evolution of the progenitor population of SNe~Ia with redshift. Studies of the SN~Ia properties and their host galaxies suggest that it is likely that SNe~Ia arise from more than one progenitor channel \citep[e.g.,][amongst others]{Scalzo2014,Childress2015,Maguire2018,Dhawan2018,Flors2020}. It is, therefore, also possible that the relative fraction of the different channels giving rise to SNe~Ia is not the same in the local and high-$z$ universe leading to a redshift evolution of the SN~Ia brightness, that cannot be corrected by the current standardisation process. This would limit our ability to disentangle this evolution from differences in dark energy properties \citep[e.g.][]{Riess2006}. 
\begin{figure*}
\centering
	\includegraphics[width=19cm]{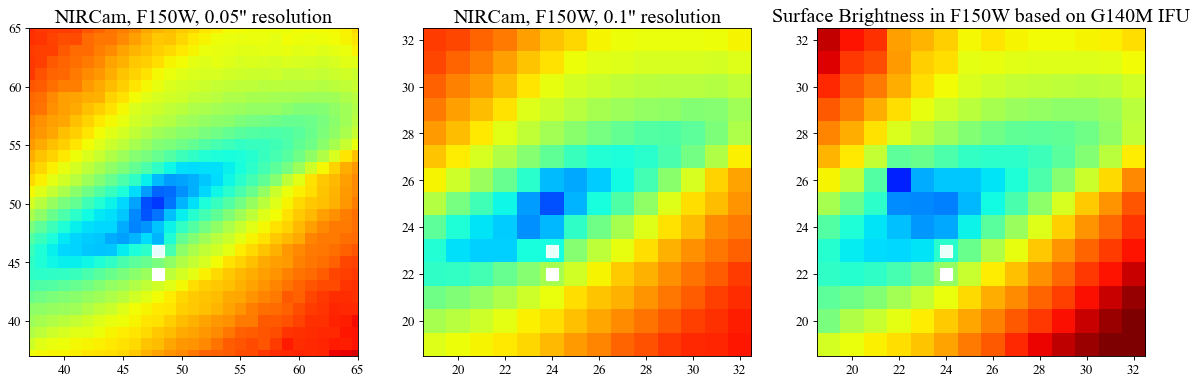}
    \caption{A summary of the NIRCam imaging and NIRSpec G140M IFU data obtained for SN~Encore (corresponding to the brightest image, i.e. Image A). Left: F150W surface brightness map calculated using NIRCam images with 0.05$^"$ resolution with a North up orientation.  Bluer color indicates brighter pixels. White dots indicate the supernova positions covered by the spaxels selected from the IFU datacube. The supernova is distinct in the image compared to surrounding pixels with lower surface brightness. Middle: NIRCam image with 0.1 $^"$ resolution with the same orientation.  We mark the pixels of potential supernova location as white dots. Right: F150W surface brightness map from the G140M IFU cube with the predicted supernova location in white.}
    \label{fig:reduction}
\end{figure*}

Strong gravitational lensing is a novel window to studying high-$z$ SNe. The ``gravitational telescope" effect \citep{Zwicky1937} magnifies the light from distant sources, allowing us to study the systems in detail. Spectra of gravitationally lensed SNe (gLSNe) have been used to study SNe~Ia near maximum brightness in the ultra-violet wavelengths \citep{Petrushevska2017} and with a time-series to study optical spectral feature evolution \citep{Johansson2021a,Gall2024}. gLSNe have also been proposed \citep{Refsdal1964,Goobar2002} as an independent way of measuring $H_0$ via time-delay distances, demonstrated with the observations of SN~Refsdal \citep{Kelly2023}, and of SN H0pe \citep{Frye2023,Pascale2024,Pierel2024a,Chen2024}. As SNe~Ia play a central role both in the study of dark energy properties and galactic chemical enrichment, it is important to have a detailed view of the explosions at $z >1$. At intermediate redshifts ($z \sim 0.5$), there have been studies comparing the spectroscopic properties, both the optical and the ultraviolet (UV) to low-$z$  ($z < 0.1$) SNe \citep[e.g. comparisons in ][]{Sullivan2009,Foley2012,Maguire2012}. It is well known that the post-standardisation SN~Ia luminosity depends on the host galaxy properties \citep{Hamuy1996,Sullivan2003,Sullivan2010,Kelly2010}. Therefore, since the galaxy properties (mass, age, metal and dust content) change with cosmic time, it is clearly important to quantify how the SN~Ia properties differ with redshift. \citet{Hook2005} analysed a sample of 14 SNe~Ia in the redshift range $0.17 \leq z \leq 0.83$, finding the spectroscopic time series and the Ca II H\&K velocities to be indistinguishable from low-$z$ SNe~Ia. Using a large sample of SNe~Ia from the Supernova Legacy Survey up to $z \sim 1$ \citet{Bronder2008} found no evolution in the rest-frame velocities and pseudo equivalent widths of the SNe~Ia. \citet{Sullivan2009} find - by comparing low-$z$ spectra to the intermediate- and high-$z$ SNe \citep{Ellis2008, Riess2007} that changes in the average spectroscopic properties are primarily attributed to the evolution in the SN~Ia demographics with redshift. Accounting for the different stretch distributions across redshift, they do not find differences between the low-$z$ and high-$z$ SNe~Ia. \citet{Maguire2012} compared a sample of low-$z$ near-UV spectra to the high-$z$ sample from \citet{Ellis2008} and found a UV flux deficit in the distant SNe~Ia at the $\sim 3\sigma$ level. \citet{Balland2009} found that $z > 0.5$ had shallower features of intermediate mass elements compared to the $z < 0.5$ sample. Seen synoptically, the studies would suggest a small, potentially appreciable difference but such an effect is far from certain.  The studies present very high-fidelity spectra but they are limited in the phase range (typically $< 10$\,d from maximum light) and are mostly at $z < 1$. It is therefore critical to study the spectroscopic properties of SNe~Ia at larger lookback times. Most spectra of intermediate redshift SNe~Ia are also observed near maximum light, therefore, it is important to study the spectral energy distribution (SED) at different phases of the explosion. 

Here, we study the post-maximum, rest-frame optical spectrum of a lensed SN~Ia, named SN~Encore, at $z = 1.95$. SN~Encore was discovered on 2023 November 17 by the James Webb Space Telescope (JWST-GO-2345, PI Newman). The discovery epoch includes spectroscopic and photometric observations, details of which are summarised in \citet{Pierel2024b}. Remarkably, the host galaxy of SN~Encore also hosted SN~Requiem \citep{Rodney2021}, which was a photometrically classified SN~Ia. The data for SN~Encore allows us to compare the properties of SNe~Ia at a lookback time of 10.5 Gyr, i.e. when the universe was around 3.3 Gyr old with the properties of SNe~Ia at present day. The observations of SN~Encore were obtained at $\sim$ few restframe weeks after maximum. At these epochs, the ejecta temperature has decreased to $\sim 7000$\,K such that there is a recombination wave that leads to an ionisation transition from doubly to singly ionised species of iron group elements \citep{Kasen2006,Blondin2015}.  This manifests as a rebrightening in the redder filters, seen as a second maximum in the  $izYJHK$ filters \citep[e.g.,][]{Hamuy1996,Folatelli2010,Dhawan2015} and a shoulder in the $Vr$ filters. The timing of this second peak can be used to measure the total amount of radioactive material in the ejecta \citep{Kasen2006,Dhawan2016}.  Along with intermediate mass elements, e.g. Si, Mg, S, at these phases, the spectrum also has lines from iron group elements \citep{Jack2015}. Such spectroscopic studies are important to test what the explosion scenario and progenitor of an SN~Ia is. While it has been known that SNe~Ia arise from the thermonuclear explosion of a carbon-oxygen (CO) white dwarf (WD) in a binary system \citep{Livio2018}, the nature of the progenitor, e.g. the mass of the WD and the companion star is poorly understood. Potential suggestions include sub-Chandrasekhar mass ($M_{\rm Ch}$) WD which detonates from a secondary supersonic shock triggered by a Helium shell detonation \citep{Woosley1994,Shen2021}. An alternate scenario is a sub-M$_{\rm Ch}$ double detonation of a hybrid He-CO WD \citep[e.g.,][]{Perets+19,Pakmor+21+hybrid,Pakmor2022,Roy+22_DDT}. Comparing the model predictions to observations of high-$z$ SNe~Ia is an important route to understanding the progenitor channels at earlier epochs in cosmic history. Such comparisons are possible  due to  suite of new photometric and spectroscopic instruments with the James Webb Space Telescope (JWST), e.g. NIRCam \citep{Rieke2005} and NIRSpec \citep{Jakobsen2022}.% Alternate progenitor scenarios  \sdt{some connections to models} 
\vspace{-0.001cm}
In this paper, we present and analyse spectroscopic observations of an SN~Ia at $z \sim 2$ coinciding with the phases corresponding to the second maximum, some of the first at this phase and high-$z$ \citep[see also][]{Pierel2024b}.  Therefore, we infer the spectroscopic properties of SN~Encore and compare then to low- and intermediate- redshift spectra from the literature to test for any evolution of SN~Ia properties with cosmic time. 
The dataset is presented in section~\ref{sec:data}, results in section~\ref{sec:results}. Results are discussed and conclusions presented in section~\ref{sec:discussion}.

%\begin{figure}
%    \centering
%    \includegraphics[width=.5\textwidth]{figures/Encore_lowzspec_comp.pdf}
%    \vspace{-0.3cm}
%    \caption{Comparison for the G235M (violet) and G140M (black) spectra with low-$z$ SN~Ia spectra in the phase range between +20 and +30\,d from peak in the restframe.    \sdt{think of two things (a) do we need the plot? (b) if yes, different SNe/ spectra for comparison?}}
%    \label{fig:spec_plot}
%\end{figure}

%\begin{figure}
%    \centering
%    \includegraphics[width=.48\textwidth]{figures/age_cons_uncons_maxrlap.pdf}
%    \caption{Comparison of the relative ages from fitting SNID to both spectra and computing the age difference (blue) and using only the same SNe that match both spectra (orange)}
%    \label{fig:snid_diff_uncorr}
%\end{figure}

\section{Data}
\label{sec:data}
SN Encore was discovered in the F150W NIRCam imaging taken 2023
November 17 (MJD 60265) by noticing a point source on top of a smooth galaxy distribution and then comparing the data with an  HST WFC3/IR F160W image
taken on 2016 July 18, MJD 57587 \citep{Newman2018a}. The filters are well-matched in wavelength and transmission and the source was brighter than 24 AB mag. However, the separation of the source from the nucleus of the host galaxy, MRG-M0138,  is $\sim 0.1^{\arcsec}$, which is approximately the pixel scale of WFC3/IR.    Forced photometry confirmed that there was an increase in flux corresponding to the apparent brightness of SN Encore between the F160W and F150W imaging. Two images of the SN: A and B were clearly visible in the F150W data, while a third image C was revealed after subtracting the host galaxy light \citep{Pierel2024b}. Lens model predictions suggest that the next image will arrive by the end of the decade, allowing a dedicated follow-up campaign similar to SN~Refsdal \citep{Kelly2016a}. The host galaxy of the SN - MRG-M0138 is red and passive, already suggesting that there is low star formation, and hence, that the SN is likely of Type Ia \citep{Rodney2021}.

The data used include {\it JWST} NIRCam images and NIRSpec IFU spectroscopic observations, both from a Cycle 1 program (Program ID: 2345, PI: Andrew B. Newman). Observations of G235M were taken on 2023 November 17, and G140M data were taken on 2023 December 27. We used the following combination of grating and disperser: G140M/F100LP and G235M/F170LP, with total exposure times of 8228s and 7644s, respectively. Exposures with the same integration time were also taken on a nearby background field. Data are processed by a modification of the JWST pipeline with version 1.13.4. Raw data were downloaded from The Barbara A. Mikulski Archive for Space Telescopes (MAST). For this work we used the {\it F150W} NIRCam images.  For NIRCam image reduction the CRDS reference files were defined by {\tt jwst\_1230.pmap}. In stage 1, the snowball rejection was activated. In stage 2, we used the default parameters in the pipeline. In stage 3, the final output image was oriented to be North-up by customizing the {\tt rotation} parameter in the resampling step, and the {\tt crval} parameter was modified to specify the 
right ascension (RA) and declination (Dec) of the reference pixel for alignment with observations made by different instruments. Two output images with different spatial resolutions, 0.05$^{\arcsec}$ and 0.1$^{\arcsec}$ per pixel, were generated.
For NIRSpec IFU data we processed G140M/F100LP 
and G235M/F170LP spectra.  The CRDS reference files are defined by {\tt jwst\_1225.pmap}. Modifications included activating {\tt NSclean} and turning off the default background subtraction to post-process sky subtraction instead of using the default 1D sky subtraction. In the {\tt cube\_build} step of stage 3, the output was oriented to be North-up, and parameters {\tt ra\_center}, {\tt dec\_center}, and {\tt cube\_pa} were specified.
\begin{figure*}
    \centering
\includegraphics[width=.85\textwidth]{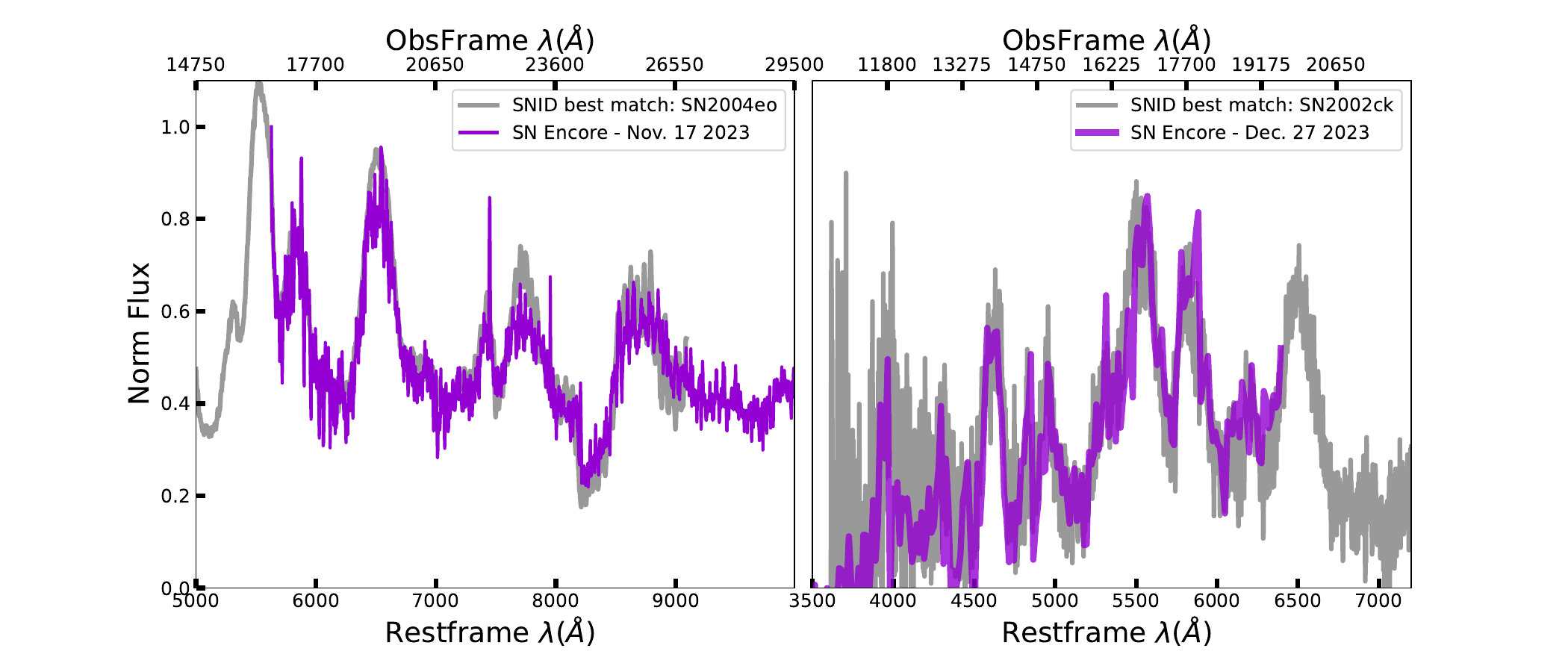}
    \caption{SNID fit to the spectrum taken on 2023-11-17 with the G235M grating and 2023-12-28 with the G140M grating. They were fitted independently, both with the host redshift free and fixed (see text for details). For the first epoch the top match is SN~2004eo at +29d and eight of the top 10 matches is ``Ia-norm". For the second epoch, the best match is SN~2002ck, a normal SN~Ia at +37\,d, also with eight out of the top 10 matches being normal SN~Ia.}
    \label{fig:snid}
\end{figure*}
\vspace{-0.01cm}
The supernova is visible in the high-resolution (0.05$^"$/ pixel) NIRCam image (left panel of Figure~\ref{fig:reduction}) but less distinct in the 0.1$^"$ / pixel NIRCam image (middle panel of Figure~\ref{fig:reduction}) and NIRSpec IFU data (right panel of Figure~\ref{fig:reduction}). We first aligned the spectra and images by specifying coordinate-related parameters and ensuring the rotation angle was consistent with North-up. Surface brightness maps were generated using filter {\it F150W}, and the IFU center was adjusted to minimize the difference between the surface brightness map from the NIRCam image and the NIRSpec IFU G140M/F100LP cube.
We performed pixel-to-pixel subtraction of background spectra from the target datacube after cleaning outliers from the background. Host galaxy properties for MRG-M0138 are described in \citet{Newman2018a}. We isolated the supernova spectra by subtracting an estimated host galaxy spectrum at the same surface brightness, using both the G140M and G235M gratings. The G140M grating enabled alignment with the F150W NIRCam image, revealing the SN's location. G235M spectroscopy was aligned with G140M using their overlapping wavelength ranges. SN spaxels showed unique spectral features distinct from the galaxy's spectra.  As shown in Figure~\ref{fig:reduction}, we selected two spaxels in the IFU data as the potential locations of the supernova.
\begin{figure}
    \centering
    \includegraphics[width=.48\textwidth]{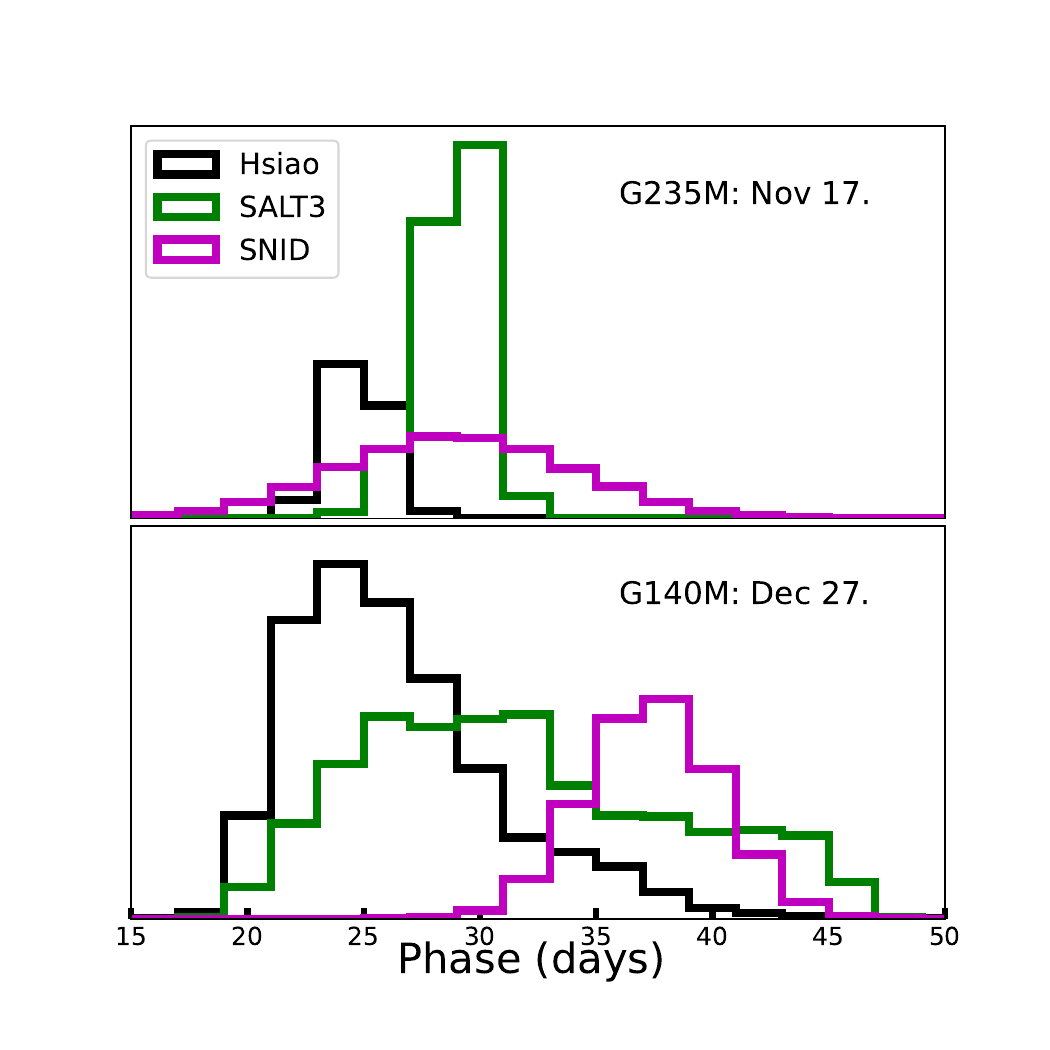}
    \vspace{-0.5cm}
    \caption{Posterior distribution of the inferred age for the G235M spectrum (top) and the G140M (bottom) for SN~Encore using the three methods, i.e. \texttt{hsiao} template (black), SALT3 (green) and \texttt{SNID} (magenta). The methods yield ages which are consistent with each other albeit with large errors.}
    \label{fig:template_age}
\end{figure}
\begin{figure*}
    \centering
   \includegraphics[width=.48\textwidth, trim=0 20 0 30]{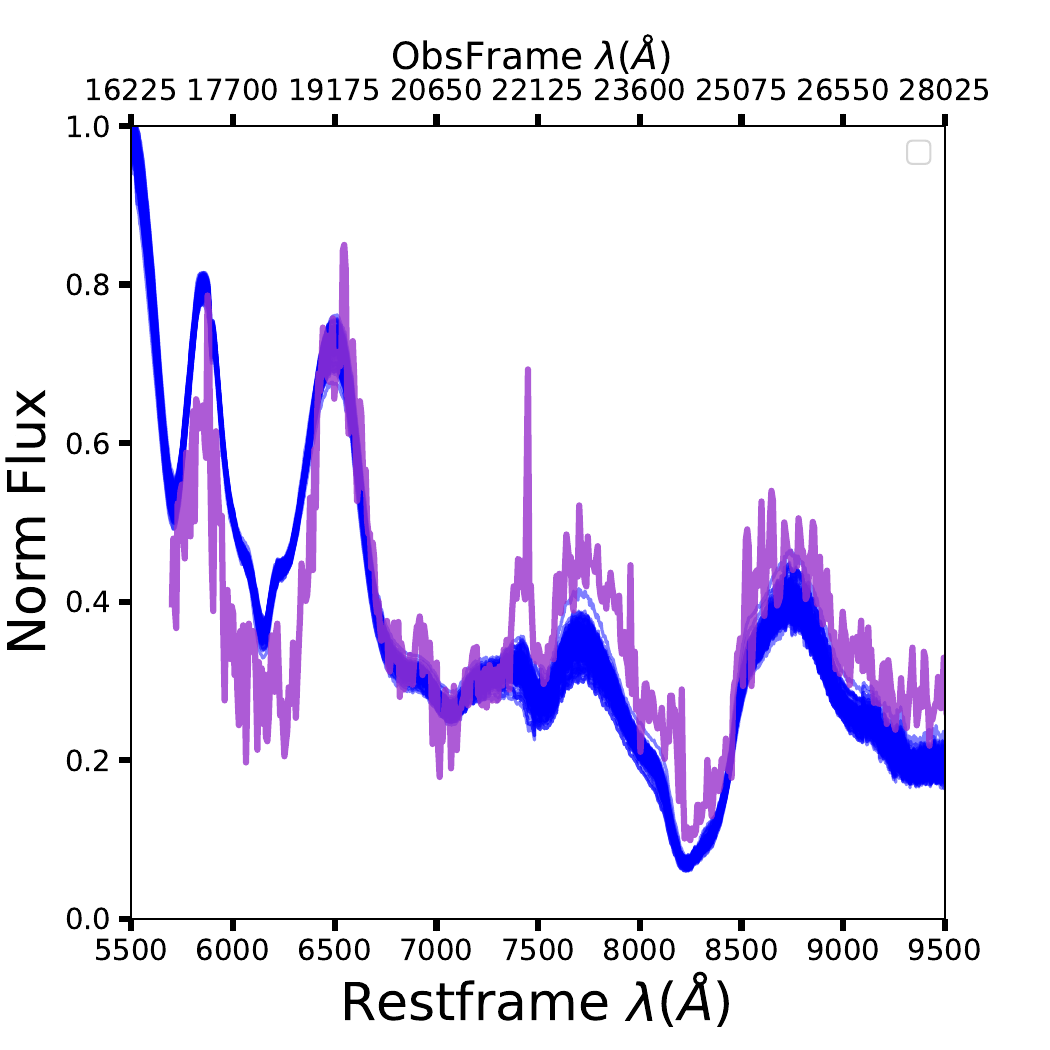}
   \includegraphics[width=.48\textwidth, trim = 0 20 0 30]{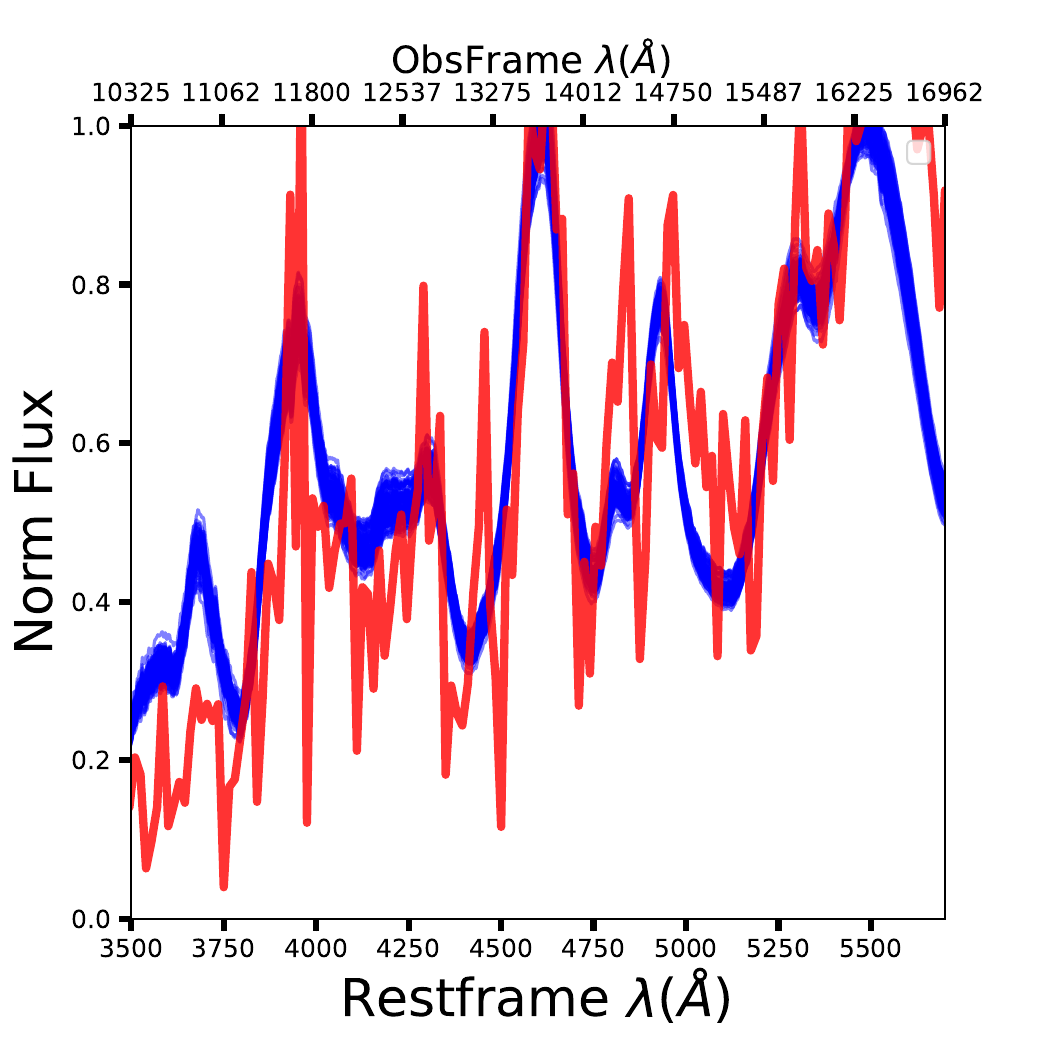}
    \caption{(Left) Comparison of the mean optical spectrum from the \texttt{kaepora} database in the phase range +17 to + 45d (blue) with the SN~Encore spectrum taken on Nov. 18 2023 (violet). The spread in the mean spectrum is from the 500 bootstrap realisations of a composite spectrum from a total of 321  spectra in the phase range without telluric contamination. Since SN~Encore is a ``Ia-norm" with a close to normal $\Delta m_{15}$ we restricted the range of comparison objects to $0.8 < \Delta m_{15} < 1.5$. (Right) The same but for the December 27 2023 (red) compared to the composite spectrum from the SNe~Ia between phase ranges +17 to +45d. }
    \label{fig:spec_encore}
\end{figure*}
\vspace{-0.01cm}
For host galaxy subtraction, we established contour levels based on the F150W surface brightness map for the G140M datacube, which allowed us to extract the background galaxy's spectra from spaxels matching the surface brightness levels of the supernova spaxels. Two spaxels for supernova locations are marked as white dots in Figure~\ref{fig:reduction}. Neighboring pixels from the supernova spaxels were masked when calculating host galaxy spectra. The host spectra were calculated as the median spectrum of the host galaxy in the same value levels of the two supernova spaxels, which include 20 and 46 spaxels for the two supernova spaxels, respectively. Finally, the SN spectrum was isolated by subtracting the median spectrum of the background galaxy from the observed SN spaxels. The extracted spectra are shown in figure~\ref{fig:snid}.
%As part of the JWST-GO-2345 program, SN~Encore was observed with NIRSpec IFU spectroscopy on November 17, 2023. While the G235M grating was observed on the day, due to a failure mode on the G140M grating, observations were taken 39 observer frame days later on December 27, 2023. 
%SN spaxels have unique features in the G235M observations, allowing us to disentangle them from the galaxy spectra.  We establish contour levels for each IFU datacube by calculating the median flux across the entire
%wavelength range, which allows us to extract the
%host galaxy spectra from the spaxels %containing
%SN flux. We establish contour levels for each IFU datacube by calculating the median flux across the entire wavelength range, which allows us to subtract the
%host galaxy spectra from the spaxels containing SN flux. Finally, we isolate the SN spectrum by subtracting off the median spectrum of host galaxy from what is observed in the SN spaxels. \sdt{add Meng's description on reduction}

\section{Results}
\label{sec:results}

In this section, we infer the SN type and age of SN~Encore spectroscopically.  We infer the phase of the observations from different spectral matching procedures. Firstly, we use the supernova identification code \texttt{SNID} \citep{Blondin2007} to simultaneously infer the SN type and age. Secondly, we use empirical spectral templates, namely, the Hsiao template \citep{Hsiao2007}
 and the latest iteration of the Spectral Adaptive Lightcurve Template - SALT3 \citep[]{Kenworthy2021} to infer the spectral age. We then test for signatures of evolution with cosmic time, we compare the SN-Encore spectrum to an ensemble spectrum from local universe ($z < 0.1$) SNe~Ia, corresponding to the phase range around the same time as the SN~Encore spectra.
 We compare the SN~Encore spectra to composite spectra of normal SNe~Ia using the \texttt{kaepora} database \citep{Siebert2019}, described in section~\ref{ssec:composite_spec}. 
 %\texttt{kaepora} is a relational database that homogenises the diverse spectroscopic observations of low-$z$ SNe~Ia, corrects them for extinction both in the Milky Way and the host galaxy and remove objects with telluric contamination (see details below and in \citet{Siebert2019}). To enable a direct comparison we only use spectra in the phase range indicated from the \texttt{SNID} fits. 
We also quantify the similarity by comparing the spectral line velocities from individual low-$z$ SNe~Ia with those of SN~Encore.

\subsection{Typing and Age inference}
We infer the type and age of the SN by fitting \texttt{SNID} to the observed spectrum. We infer the spectral properties in the first instance with no prior on the fit. We find a best fit to a normal Type Ia supernova (Ia-norm), with the best fit to SN~2004eo. Nine of the ten top matches are to ``Ia-norm" SNe.  We independently fit the second epoch to get the best SNID match. Eight of the top 10 matches for the spectrum taken on 2023-12-27 are also to SN~Ia-norm with a best fit to SN~2002ck (Figure~\ref{fig:snid}).

The age inference is based on the top 8 fits with rlap, a metric used by SNID to determine how correlated/similar the two SN spectra are,  $> 5$ and grade classified as``good". We take the median and standard deviation of the top 8 fits \citep[as demonstrated in][]{Blondin2008} as the inferred phase and find that the spectrum suggests a phase of 27.9 $\pm 5.3$ days. In the second instance, we fit with a prior on the redshift from the host galaxy. If we fix the redshift in the fit to that from the host galaxy spectrum, we get a consistent estimate for the phase of 29.0 $\pm 5.0$\,d. We use this estimate for the analysis from here onwards. For the second epoch, fitting independently of the first spectrum,
the best estimate of the age using the same methods as for the first epoch is 37.4 $\pm 2.8$\,d. For this epoch it is independent of the prior on the redshift from the host galaxy.
Therefore, based on the spectroscopic information we conclude tha SN~Encore a normal SN~Ia. 
%\subsection{Comparison to intermediate-$z$ SNe~Ia}
%\subsubsection{Spectral Template Age Inference}

We use a complementary technique to measure the phase of the spectra, with empirical templates for the SED. Since there are few spectral templates, we compare the results from different model fits. The models are fitted using a Markov Chain Monte Carlo (MCMC) with \texttt{emcee} \citep{Foreman-mackey2013} with 20000 steps and 5000 samples for ``burn-in".  In the first case, we use the Hsiao template \citealt{Hsiao2007}, which is built from a library of $\sim 600$ spectra of $\sim 100$ unique SNe~Ia. Fitting this template to the first observed spectrum of SN~Encore we find a best matching phase of 24.7 $\pm 1.1$\,d. 
When using the latest SALT3 model \citep{Kenworthy2021}, which was developed from $\sim$ 1200 spectra of 380 distinct SNe Ia, we get a phase of 28.7 $^{+2.0}_{-1.5}$ \,d which are in agreement at the $\sim 1 \sigma$ level. Similarly for the G140M spectrum, the Hsiao model indicates an age of 25.6$^{+3.1}_{-4.5}$\,d and SALT3 model indicates an age of 30.9 $^{+4.4}_{-9.1}$\,d. All the methods yield consistent phase inference to within 2 $\sigma$ (Figure~\ref{fig:template_age}). The difference between the phases inferred for both spectra are consistent with the difference in the dates of observation, accounting for time-dilation.

\subsection{Low-$z$ spectral comparison}
\label{ssec:composite_spec}

\begin{figure}
    \centering
    \includegraphics[width=.5\textwidth]{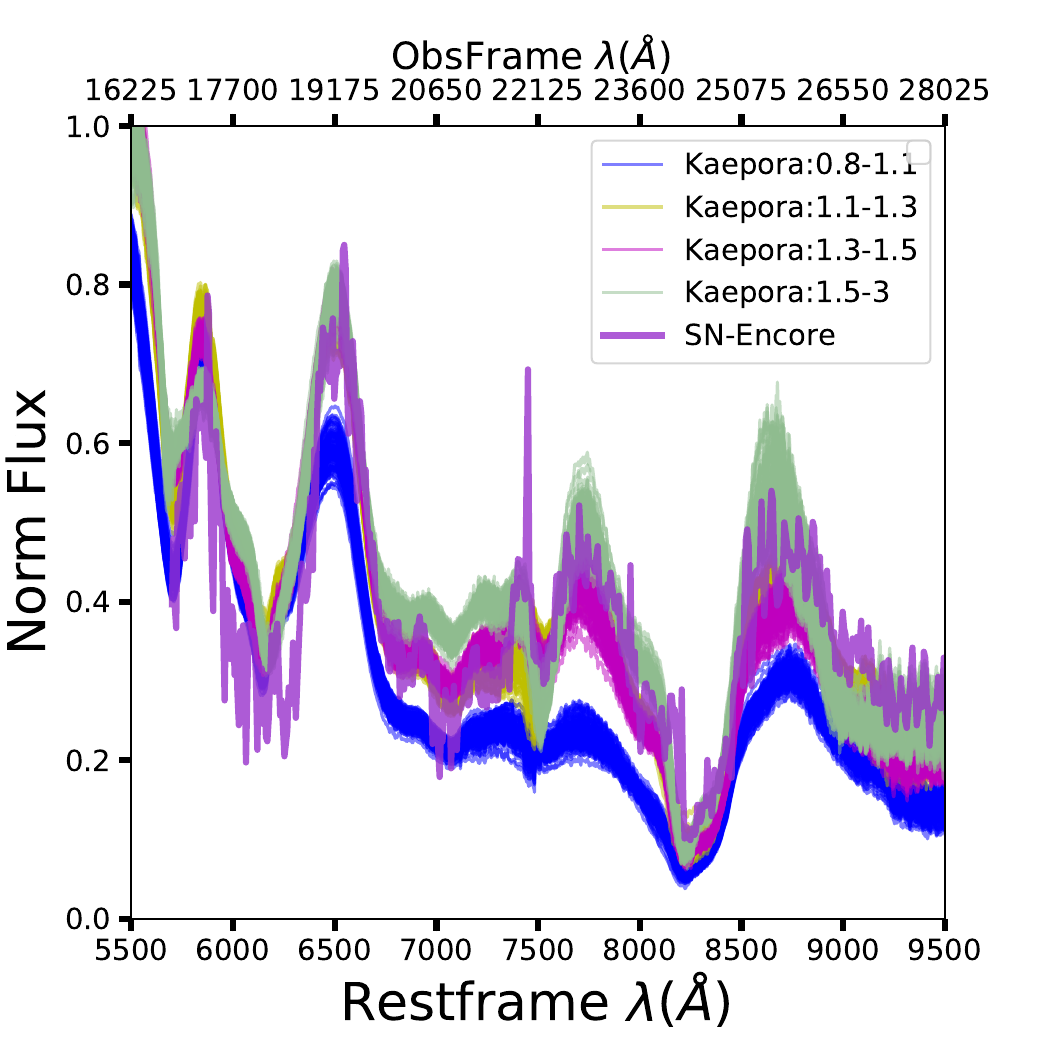}
    \caption{Comparison of the mean low-$z$ SN~Ia spectrum in the phase range permissible from the \texttt{SNID} inference, binned in $\Delta m_{15}$. All the composite spectrum with a median $\Delta m_{15} > 1.1$ are consistent with the first observed spectrum of SN~Encore. }
    \label{fig:dm15_split_kaepora}
\end{figure}

An important test of the evolution of SNe~Ia with cosmic time is the comparison of spectra between low-$z$ and high-$z$ SNe \citep[e.g.,][]{Ellis2008, Petrushevska2017}. 
We compare the SN~Encore spectra to the local SN~Ia sample using the relational database \texttt{kaepora} \citep{Siebert2019}. We restrict the low-$z$ sample to spectra between +17 and +45\,d, which is the possible range of phases for SN~Encore. \texttt{kaepora} contains a large database of almost 5000 spectra of nearly 800 SNe~Ia. The data are homogenised to account for diverse signal to noise ratios, cleaned by removing residual sky lines, galaxy emission and cosmic rays and corrected for host and Milky Way extinction as well as deredshifted to the restframe of the SN. 
We make composite spectra from the database to compare with SN~Encore (Figure~\ref{fig:spec_encore}). We begin by removing peculiar SNe~Ia, which are defined as being similar to SN~2002cx, 2002es or classified as Ia-CSM. 
We generate 500 bootstrap realisations from the composite spectra as described in \citet{Siebert2019}. Composite spectra allow us to visualise  differences between complex spectral feature shapes, which might not be possible with summary statistics, e.g., pseudo-equivalent widths  or line velocities. The low-$z$ composite spectra are compared to the G140M and G235M spectrum of SN~Encore in Figure~\ref{fig:spec_encore}. There are no obvious differences between the low-$z$ spectra and SN~Encore in this phase range. We also compare the SN~Encore spectra to low-$z$ spectra in \dmfif\, bins for values from 0.8 to 2.0. For comparison to the inference from the photometric data, we use an $x_1 = -1.39 \pm 0.13$ (Pierel et al. in prep.), where $x_1$ is the lightcurve shape derived from the SED model used in cosmological inference, i.e. SALT3 \citep{Kenworthy2021} . The negative $x_1$ suggests that the SN is declining faster than the mean, while still being within the distribution expected for cosmological SNe~Ia. Using the relation in \citet{Guy2007}, this corresponds to a \dmfif\, of $1.34 \pm 0.02$. While we do not explicitly restrict either the comparison samples or models for our analysis, we use this inferred \dmfif\, as an indicative value. A comparison of the composite spectra in bins of \dmfif\, is shown in Figure~\ref{fig:dm15_split_kaepora}. We find that the composite spectra for the slowest declining SNe~Ia are the least consistent with the observed spectra of SN~Encore while the other composite spectra agree very well. This is an independent test that SN~Encore is a cosmological SN~Ia with a faster than average decline rate. 
%For comparison to local SNe~Ia we use observations from the Carnegie Supernova Project \citep[CSP;][]{Folatelli2013}, a multi-year observing campaign of low-$z$ ($z \lesssim 0.1$) SNe. The spectral dataset in the optical wavelengths has objects observed at large phase ranges. The CSP dataset has XX spectra in the phase range corresponding to the SNID inferred phase of SN~Encore. 

SN~Ia observing campaigns have also obtained high-fidelity spectra at intermediate to high-$z$ \citep[e.g.][]{Ellis2008,Balland2009,Balland2018}. We note that while a comparison to composite spectra of intermediate-$z$ SNe~Ia would be very interesting, it is not viable since  the spectra are observed close to maximum light, unlike the inferred phase of both the G140M and G235M spectra.
\begin{figure}
    \centering
    %\vspace{-0.1cm}
    \includegraphics[trim =0 30 0 30, width=.48\textwidth]{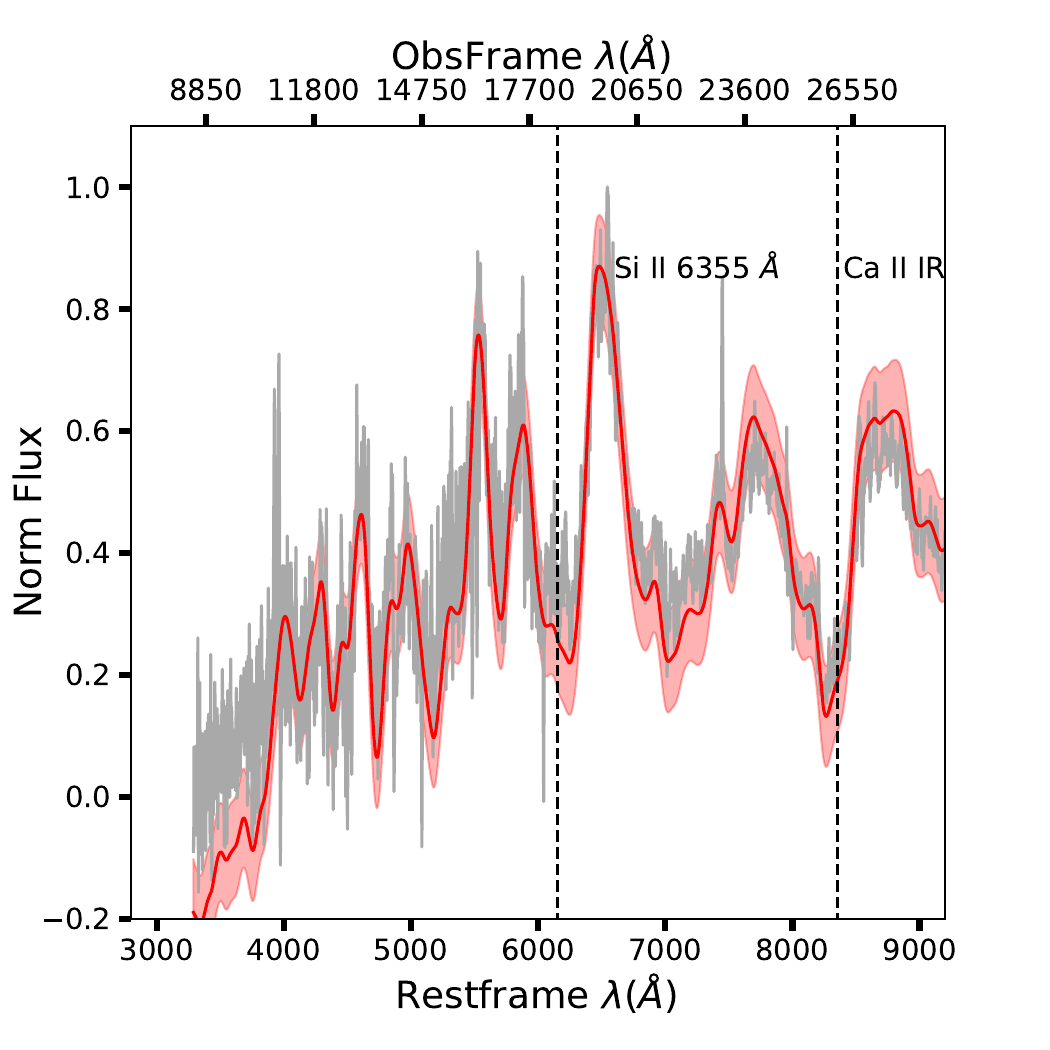}
    \caption{Gaussian Process fit from \texttt{spextractor} \citep{pap19} to the SN~Encore spectrum (joined G140M and G235M). }
    \label{fig:encore_spextract}
\end{figure}
\subsection{Spectral Feature Comparison}
To quantify the spectroscopic comparisons further, we measure the spectral line velocity  of the most ubiquitous features in an SN~Ia, i.e. the Si II 6355 $\AA$ and the Ca II NIR triplet. Spectral lines bluewards of the Si II 6355 $\AA$ feature cannot be measured very precisely with either spectra, hence, we only analyse the two aforementioned features, observed in the G235M spectrum. The bluer features are also difficult to measure without contamination at phases post maximum \citep[e.g.][only measure it to $\sim +20$ days post peak]{Folatelli2013}.  
\begin{figure*}
    \centering
\includegraphics[width=.48\textwidth]{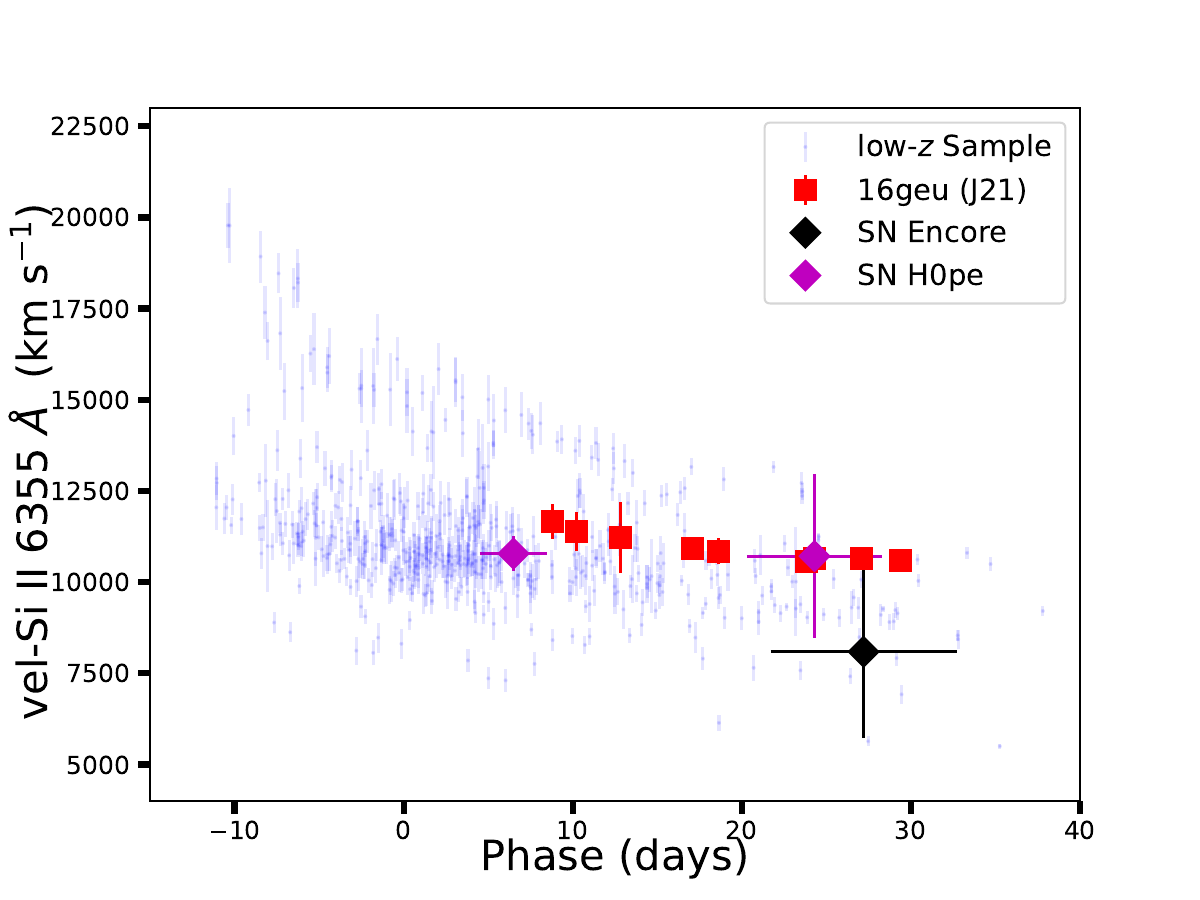}
\includegraphics[width=.48\textwidth]{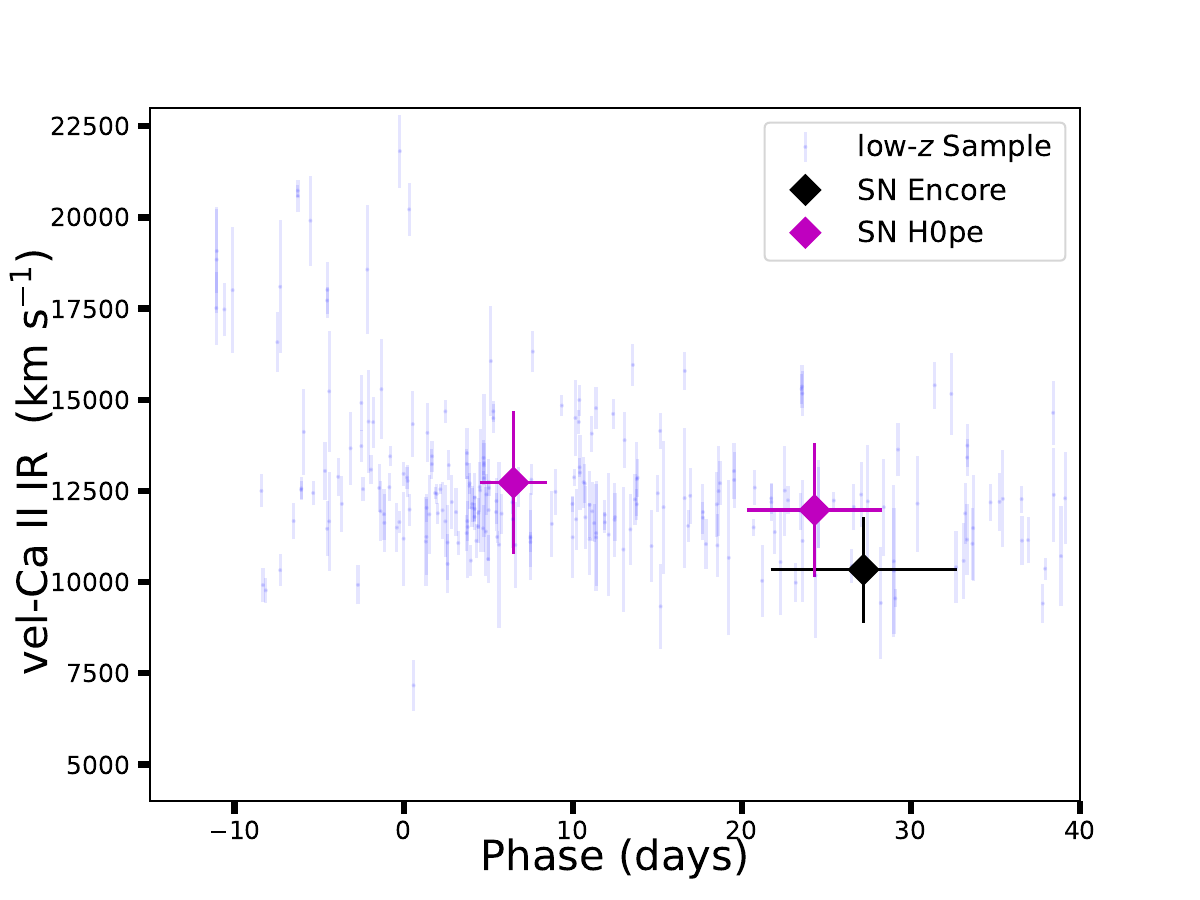}
    \caption{The evolution of the Si II 6355 $\AA$ (left) and Ca II NIR triplet line velocity (right) for low-$z$ SNe~Ia compared to lensed SNe~Ia at high-$z$.  The low-$z$ sample is shown in blue,  and SN~Encore in black. iPTF16geu \citep{Johansson2021a} is shown in red and SN H0pe \citep{Chen2024} in violet.   We find that the the velocity of SN~Encore is consistent with the low-$z$ sample for the Si II 6355 $\AA$.}
    \label{fig:vel_evol}
\end{figure*}
%\begin{figure}
%    \centering
    %\includegraphics[width=.48\textwidth]{figures/phase_salt_hsiao.pdf}
%    \caption{Inferred posterior distribution of the restframe phase of the spectrum from the Hsiao (green) and SALT3 (purple). The results are consistent between the models and with the SNID fit. Note that the plotted errors only include the statistical errors from the spectral fit.}
%    \label{fig:phase_salt_hsiao}
%\end{figure}
The spectra are  fit with a gaussian process (GP) smoothing using the tool \texttt{spextractor}\footnote{https://github.com/astrobarn/spextractor} \citep[see also,][]{pap19,Burrow2020}.  
The spectral line velocities are inferred from doppler shift between the wavelength minima $\lambda_m$ computed by \texttt{spextractor} using the GP smoothing and the rest frame wavelength of the feature (see Figure~\ref{fig:encore_spextract} for the resulting fit). %This is computed using the doppler shift formula, i.e. 
%\begin{equation}
%\centering
%    v = \frac{c \left[ \frac{\lambda_m^2}{\lambda_0}^2 - %1\right]}{\frac{\lambda_m^2}{\lambda_0}^2 + 1}
%\label{eq:vel_exp}
%\end{equation}

%The other spectral diagnostic which gives an estimate of the depth of the feature is the pseudo equivalent width (pEW). Its given by

%\begin{equation}
%    pEW = \sum_i \left(1 - \frac{f(\lambda_i)}{f_c(\lambda_i)}\right) \Delta \lambda_i
%\end{equation}
%where $f(\lambda)$ is the observed flux at the given wavelength and $f_c(\lambda)$ is the flux at the pseudo continuum at the same wavelength (since the real continuum is absent and replaced with a straight line this is referred to as pseudo). The velocity and pseudo-equivalent width are used to sub-classify SNe~Ia \citep[e.g., ][]{Branch2006,Blondin2012,Folatelli2013,Maguire2014} and also an important diagnostic of the explosion. 
%\sdt{To-add:  A description of data for SN H0pe and iPTF16geu.}
Another metric used widely in the literature for spectral comparisons is the pseudo equivalent width (pEW). It is similar to the equivalent width (EW) but since EWs require a continuum to be determined, which is not well-defined for an SN~Ia, a pseudo-continuum is defined, e.g. as a straight line fit between two local maxima \citep[e.g., see][]{Folatelli2004}.
While pEWs are an important diagnostic of the explosion \citep[e.g., ][]{Branch2006,Folatelli2013,Maguire2014}, at high-$z$, we can expect a non-negligible contamination from the host galaxy \citep[e.g.,][]{Foley2008}. Therefore, to be conservative and use a ``cleaner" diagnostic we only analyse the velocities of the two features.

For comparing to the low-$z$ SNe~Ia, we use measurements from the low-$z$ as presented in   \citet{Folatelli2013}. This is a large, uniformly measured set of values that extend out to late phases, compatible with the phases of observation of SN~Encore. The comparison to SN~Encore is presented in Figure~\ref{fig:vel_evol}. We find that both the line velocities of Si II 6355 $\AA$ and Ca II NIR are compatible with the distribution of value for the low-$z$ sample showing no sign of evolution with redshift.
We also compare the line velocity to other lensed SNe~Ia, both at intermediate  ($0.1 < z < 0.5$) and high-$z$ ($z > 1$). 
Spectra for SN~H0pe are obtained from \citet{Chen2024} and the line width and velocity is measured with the same procedure as for SN~Encore. We find no differences in the velocity evolution of either feature compared to the low-$z$ sample for SN H0pe. For comparison we also plot the inferred line velocities for iPTF16geu, an intermediate redshift lensed SN~Ia at $z \sim 0.409$ \citep{Goobar2017}, with the line velocities as reported in \citet{Johansson2021a} where the authors found that iPTF16geu showed a velocity evolution very similar to normal SNe~Ia at low-$z$ \citep[see also][]{Cano2018}. Analysis of SN~Zwicky \citep{Goobar2023}, an extremely compact lensed SN~Ia will be presented in a forthcoming paper (Johansson et al. in prep.) 
%The comparison of the velocity and pEW is shown in Figure~\ref{eq:vel_exp}
%We also compare the pEW and line velocity evolution to other intermediate ($0.1 < z < 0.5$) and high ($z > 1$) redshift Type Ia supernovae observed at a similar epoch to SN~Encore. Since SNe~Ia typically fade by a few magnitudes at these phases compared to maximum light, the only objects with high fidelity observation at such phases are the lensed SNe~Ia, iPTF16geu \citep{Johansson2021a} and SN~H0pe \citep{Chen2024}. 

\subsection{Comparison to explosion models}
Spectra of SNe~Ia at a few weeks post maximum contain important information regarding the elemental abundance and ionisation state of the ejecta. At these phases, the IGEs in the ejecta transition from doubly to singly ionised, which is seen as a second maximum in the lightcurve at redder wavelengths (typically, $i$-band and redwards). This can be used as a diagnostic to distinguish between different model scenarios. In this section we compare the observations of SN~Encore with synthetic spectra from 
different explosion scenarios. One of the open questions regarding SN~Ia progenitors is the mass of the primary WD.  Therefore, we compare the observations to both models for which the primary WD is Chandrasekhar mass \citep[$M_{\rm Ch} \sim 1.4 M_{\rm \odot}$, e.g.][]{Khokhlov1991} or significantly below it  \citep[sub-$M_{\rm ch}$, e.g.][]{Blondin2017}. Here, we take a representative set of explosion models to cover the range of observables seen in the sample of ``normal" SN~Ia, since SN~Encore is a normal SN~Ia. We briefly describe below the different scenarios considered in this work.

For the $M_{\rm ch}$ case, the leading model is the ``delayed-detonation" scenario \citep{Khokhlov1991,Seitenzahl2013}. In this model, the explosion begins as a subsonic shock, or ``deflagration", which at a certain transition density, ``$\rho_{\rm DDT}$" becomes a supersonic shock or a detonation. Without the pre-expansion or deflagration phase, the material of the WD would completely burn to iron-group elements (IGEs) without any intermediate mass elements (IMEs), hence, the resulting explosion would not look like an SN~Ia \citep{Arnett1969}. The subsonic shock pre-expands the white dwarf reducing the density of the material, and this leads to the creation of both IGEs and IMEs giving rise to near maximum light spectra that agree well with observations \citep{Khokhlov1991}. The variation in the nucleosynthetic yields and therefore, synthetic observables is determined by the transition density, $\rho_{\rm DDT}$. These models are referred to hereafter as DDC \citep{Blondin2013}. A variation of the $M_{\rm Ch}$ delayed detonation models are the pulsational delayed detonations, which are very similar to the DDC models, except at early times wherein they have little mass at high velocity and hotter outer ejecta \citep[hereafter, PDDEL;][]{Dessart2014}. 

For the sub-Chandrasekhar (sub-M$_{\rm Ch}$) models, the mass of the primary varies, which leads to differences in the elemental yields and the abundance of radioactive material (hereafter SCH). The models can explain a wide range of brightnesses and lightcurve shapes, and there is observational evidence that faint SNe~Ia arise from sub-$M_{\rm Ch}$ explosions \citep{Blondin2017}. While previously the models were disfavoured as the outer He-shell (which was assumed to be very thick) produced Ti-group elements that predicted colours mismatched with observations \citep{Woosley1994,Hoeflich1996}, with recent thin shell models \citep[e.g.,][]{Bildsten2007,Shen2014}, the agreement with observed spectra and colours is better \citep{Shen2021a,Shen2021b, Pakmor+21+hybrid,Pakmor2022}. 

\begin{figure}
    \centering    \includegraphics[width=.5\textwidth]{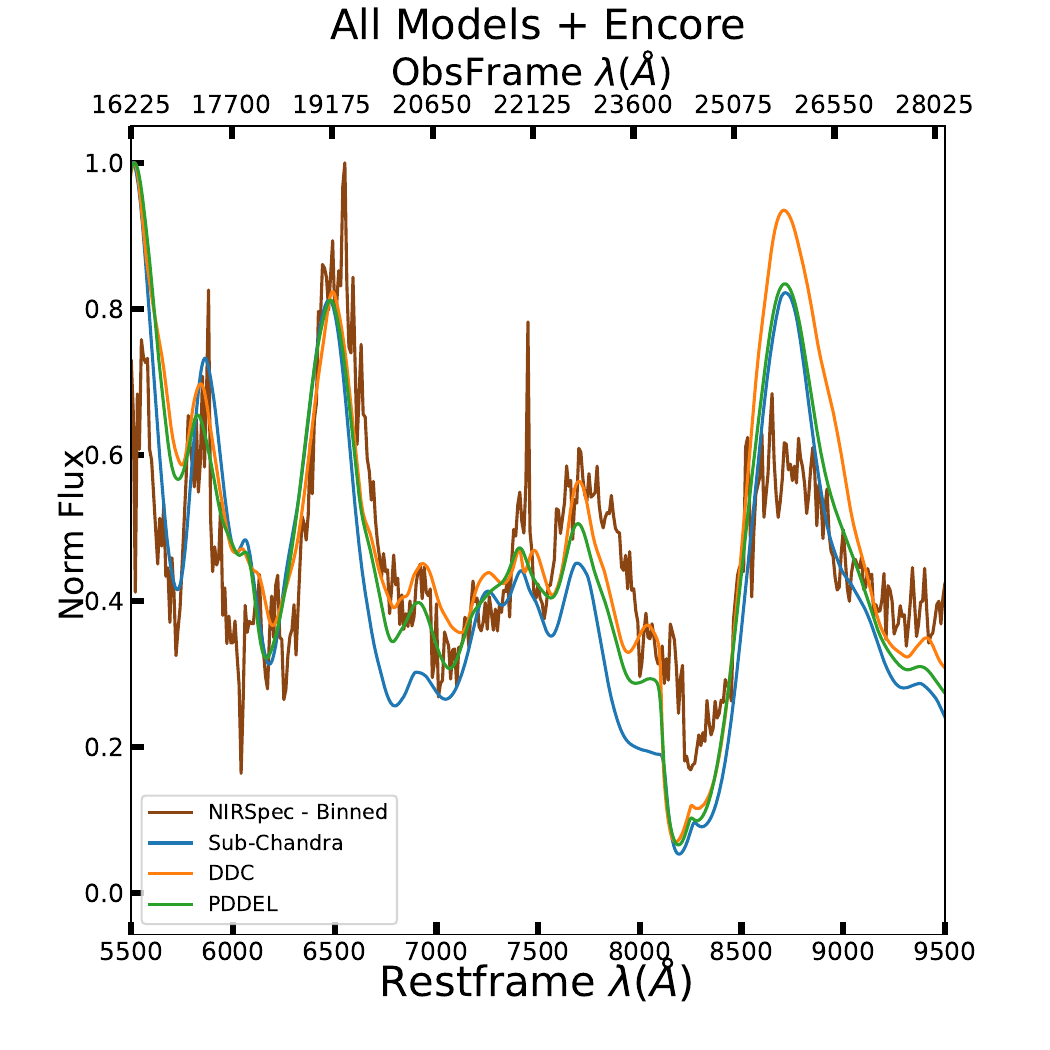}
    \caption{Comparison of the synthetic spectra for different explosion models with the observed spectrum of SN~Encore  within the estimated range based on both the photometry and the spectra.}
    \label{fig:model_bestdm15}
\end{figure}
Recent studies have shown that the population of SNe~Ia could comprise of objects arising from either progenitor channel \citep[amongst others]{Blondin2017,Dhawan2018,Scalzo2019}. Therefore, we compare the synthetic observables from all the model categories describe above to the spectra of SN~Encore.  We investigate the radiative transfer from a wide range of explosion models, such that we can cover the known diversity of progenitor scenarios.
Both the $M_{\rm Ch}$  (DDC, PDDEL) and sub-$M_{\rm ch}$ models span a range of $^{56}$Ni masses and in case of the SCH models, they also span a range of primary white dwarf masses, since in this scenario the WD mass is the main determinant of the $^{56}$Ni mass \citep[e.g.,][]{Sim2010}. 
%The DDC, PDDEL and SCH models span a range of $^{56}$Ni masses as well as other physical parameters (e.g. in case of SCH models, the mass of the primary white dwarfs). We start by comparing the observations with the theoretical spectral time series that is closest to a $\Delta m_{15}$ of 1.1, i.e. a normal SN~Ia. The DDC15, PDDEL7 and SCH4p5 models meet this criterion. 

We pick the models in each class with \dmfif\, closest to the observations of SN~Encore. For the scenarios tested, this corresponds to SCH3p5, DDC20 and PDDEL9 (figure~\ref{fig:model_bestdm15} the models are shown normalised to the flux around $\sim 6200$ \AA). We pick the model predictions closest to the phase of the G235M spectrum and find that all models can broadly reproduce the features in the G235M spectrum. Some of the most striking differences are in the region $\sim 9000$ \AA, which is likely a combination of strong Ca II, Fe II and Co II lines \citep{Blondin2015}. While all models overpredict the relative flux in both these regions, the DDC models predict the brightest line ratio. However, the predicted strength of the line is  sensitive to the phase inference, which has a few day uncertainty in our analysis. At bluer wavelengths we see that the sub-Chandra models underpredict the flux at $\sim 6700 \AA$ which could be due to differences in the Si II line strength. The DDC and PDDEL overpredict the flux at $\sim 5500 \AA$ which is likely dominated by permitted transitions of Fe II and possibly forbidden [Co III], since the SN is transitioning to the nebular phase at such late epochs. 
If the true phase is not at the central value but within either the 68$\%$ or 95$\%$ credible region, then the prediction from the DDC and PDDEL models matches the observed features better than the SCH model. Therefore the inference regarding the model which most closely resembles the spectrum depends on having a precise, independent estimate of the epoch of the observation. Moreover, we note that this comparison is only performed for a single SN. In the future, such comparisons for a sample of high-$z$ lensed SNe will be necessary to derive properties of the populations of SNe~Ia, as done with composite UV spectra at $z \sim 0.5$ in \citet{Walker2012}.  %\sdt{rewrite with only one phase}

\section{Discussion and Conclusion}
\label{sec:discussion} 
%\sdt{make sure this is updated}
We have analysed the spectra of SN~Encore, a lensed Type Ia supernova at $z=1.95$ \citep{Pierel2024b}. Due to the boost in brightness from the lensing magnification, we can study a high-quality spectrum a few restframe weeks after maximum light, of an SN~Ia with a lookback time of $>$10 Gyr. 
We find that the SN is consistent with a normal SN~Ia subclassification from \texttt{SNID} template matching. The best fit phase of observations is $29.0 \pm 5.0$\,d for the G235M spectrum and $37.4 \pm 2.8$\,d for the G140M spectrum. This is consistent with the differences in observer frame days divided by $(1+z)$ \citep[see also][for an analysis of time-dilation with SN~Ia lightcurves]{White2024}. We built a composite spectrum of low-$z$ SNe~Ia using the \texttt{kaepora} relational database, we found no significant differences between the local sample and either of the SN~Encore spectra. Binning the composite spectra based on the lightcurve shape, we find that all except the slowest evolving subsample, i.e. with $0.8 < \Delta m_{15,B} < 1.1$ match the observed spectrum, which is consistent with the measurement of the shape of the lightcurve. We find no differences between the G140M observations of SN~Encore and the mean low-$z$ spectrum in the same wavelength region. In previous studies comparing the UV spectra of low- and intermediate-$z$ ($z \sim 0.5$) SNe~Ia, there has been evidence for depressed flux in the local SNe~Ia compared to more distant SNe~Ia in UV features at $2920$\AA\, and $\sim 3180$\AA. While our observations do not cover such blue features, future observations with G140M / F070LP will be critical to compare these features between local and very high-$z$ ($z > 1$) gLSNe~Ia.

As SNe~Ia are precision probes of dark energy \citep{DES2024}, it is important to quantify systematics from any evolution of intrinsic properties with the age of the universe. The tests presented here show that based on post-maximum light spectra, there is no evidence for differences between the spectral properties of SN~Encore and low-$z$ SNe~Ia. Complementary analysis of a lensed SN~Ia at $z = 1.4$ - PS1-10afx \citep{Quimby2014} - near maximum light also demonstrated a similarity between the properties of low- and intermediate-$z$ SNe~Ia \citep{Petrushevska2017}. However, it was found that for PS1-10afx, the feature around 3500 \AA\, is more prominent than the low-$z$ sample. While there isn't significant signal in that wavelength region for SN~Encore, future observations will be critical to test whether such differences are also seen at later phases.
We quantified the comparison by evaluating the spectral line velocities of SN~Encore, as well as other lensed SNe~Ia, namely, SN~H0pe and iPTF16geu. We find that the line velocities measured for the intermediate- and high-$z$ are consistent with the distribution observed for low-$z$ SNe~Ia. 
 
 With the advent of wide-field, deep surveys of the transient sky, we expect to discover large samples of lensed SNe \citep{Pierel2019, Arendse2023}. Moreover, we can expect large samples of cluster-lensed SNe~Ia to be discovered with JWST \citep{Petrushevska2018}. With a large sample of SN spectra, we can compare to predictions from different explosion models to infer what progenitor channels can explain the observed population of SNe~Ia. Here, we illustrate this point by comparing to both $M_{\rm ch}$ and sub-$M_{\rm ch}$ models. The sub-$M_{\rm Ch}$ models underpredict the features near $\sim 6700$ \AA\, whereas the DDC and PDDEL models overpredict the complex at $\sim 5500$ \AA\,, and the DDC models predict the strongest feature at $\sim 9000$ \AA. With future improvements in the model predictions we can use the spectra for a sample of SNe~Ia to distinguish between different model scenarios. Strongly lensed supernovae are, therefore, an excellent method to test cosmic evolution of SNe~Ia both for controlling systematics in dark energy inference and understanding their progenitors.%\sdt{think more...}

\section*{Acknowledgements}
Support for programs JWST GO-2345 and DD-6549 was provided by NASA through a
grant from the Space Telescope Science Institute, which is operated by the Associations
of Universities for Research in Astronomy, Incorporated, under NASA contract NAS5-26555. This paper is based on observations with the NASA/ESA  James Webb Space Telescope obtained from the Mikulski Archive for Space Telescopes at STScI. 
We thank Umut Burgaz for interesting discussions on spectral features. SD acknowledges support from a Kavli Fellowship and a Junior Research Fellowship at Lucy Cavendish College. JDRP is supported by NASA through a Einstein
Fellowship grant No. HF2-51541.001 awarded by the Space
Telescope Science Institute (STScI), which is operated by the
Association of Universities for Research in Astronomy, Inc.,
for NASA, under contract NAS5-26555. C.L. acknowledges support from the National Science Foundation Graduate Research Fellowship under Grant No. DGE-2233066. FP acknowledges support from the Spanish Ministerio de Ciencia, Innovación y Universidades (MICINN) under grant numbers PID2022-141915NB-C21. This work was supported by research grants (VIL16599, VIL54489) from VILLUM FONDEN.
RAW acknowledges support from NASA JWST Interdisciplinary Scientist grants NAG5-12460, NNX14AN10G and 80NSSC18K0200 from GSFC.

%%%%%%%%%%%%%%%%%%%%%%%%%%%%%%%%%%%%%%%%%%%%%%%%%%
\section*{Data Availability}

Data is available via publically set repositories and webpages like \texttt{wiserep}.



\bibliographystyle{mnras}
\bibliography{mnras_template} % if your bibtex file is called example.bib

%\begin{thebibliography}{99}
%\end{thebibliography}

%%%%%%%%%%%%%%%%%%%%%%%%%%%%%%%%%%%%%%%%%%%%%%%%%%

%%%%%%%%%%%%%%%%% APPENDICES %%%%%%%%%%%%%%%%%%%%%

%\appendix


%If you want to present additional material which would interrupt the flow of the main paper, it can be placed in an Appendix which appears after the list of references.

%%%%%%%%%%%%%%%%%%%%%%%%%%%%%%%%%%%%%%%%%%%%%%%%%%

% Don't change these lines
\bsp	% typesetting comment
\label{lastpage}
\end{document}